# Contrastive sets and framing: A case study in scientific writing

Paul J. Camp
*Spelman College, Department of Physics, Atlanta, GA, 30314*

I describe an activity created to help our students learn how to write a scientific paper by reverse engineering a contrastive set of existing papers. I look at three recent implementations of this activity and use a multiple cognitive models approach to begin understanding the variance. This leads to a reasonable strategy for further investigation. First, I describe the conception and design of the activity structure, and how a contrastive set was constructed. I create an initial coding scheme to characterize group behavior in the three implementations, with special attention paid to analyzing student dialog on the structure of scientific papers and their ideas about the reasons for the peculiar features they observe. There is a striking difference in the way students engaged with the activity between the implementations not all of which can be ascribed to differences in activity structures, and the remainder of the paper is concerned with the use of resources and framing, case-based reasoning and Gibson's ecological perspective to understand the reason for this difference.

## I. INTRODUCTION

*Can we do it like this? Can we just put our own little scientific stamp on it?*
*--Spelman Student*

As part of an ongoing effort to enculturate students into the norms and practices of physics, the Spelman College Physics Department requires our students in the introductory courses to engage in design and implementation of scientific investigations of real phenomena in the context of ill structured problems. The culmination of each investigation is a scientific paper written according to the professional standards of physics. Being mostly just out of high school, this is the first exposure most of these students have had to scientific text and so producing their own is a daunting task. Prior experience with reading and writing narrative text or writing lab reports is of limited help. In effect, they were trying to reproduce something they'd never seen based on a verbal description of what it looked like. To address this problem we have embedded a discussion-based "reverse engineering" activity to improve the quality of their papers, and that activity is the topic of this paper.

I describe the structure of the activity based on a contrastive set of readings followed by group discussion as well as an initial attempt at a coding scheme to characterize their performances. I compare three implementations. Two are ostensibly identical in class activities, and one is a product of an effort to streamline the class by moving the activity into outside reading and an on line discussion forum. The dramatic differences between the ways students engaged in the activity is difficult to interpret in any way other than as a particularly stark example of the impact of framing on performance, and that is currently my working hypothesis.

This paper is also structured as a contrastive set of case studies. My goal is not quantitative rigor so much as to establish a conceptual foundation that will guide future work. I address two questions: what are the emergent features of their discussions and work products that characterize the way they engage with the activity, and what plausible mechanisms might lead to the variation in their modes of engagement? I do not make any claims about a final or complete explanation. We were motivated to take a closer look at the activity by some odd and variable results and my intent in this paper is to frame a plausible hypothesis and data collection strategy that will serve as a foundation for future work.

The context of this activity is the introductory calculus-based first semester physics course. Both the lecture and the lab employ project-based instruction but implemented in different ways. The lecture uses mini-projects which are of limited scope and generally focused on a small set of ideas. In the lab, projects are much larger scale and involve coordination of a large set of theoretical, experimental and computational resources to model the phenomena under investigation. There are two projects: investigating a car crash based on the evidence that can be detected after it is over, and designing a bungee jump that is safe and fits biological constraints on acceleration. Each of these extends across multiple weeks and they form the context for everything that occurs in the lab.

Each group is engaged in a different experimental investigation relevant to the phenomenon so there is a need to communicate results at some point – no one can solve the overarching challenge without understanding what all the groups have discovered. This is the point at which writing a scientific paper happens. The activity to be discussed here is introduced in a just-in-time fashion when they are ready to write.

First, I will describe the activity in its original conception, derived from prior work on the use of contrastive sets in instruction. I will then analyze data collected from the most recent implementations of this design. In the recent past, an effort was made to streamline the activity and reduce its impact on class time. I will describe and analyze that implementation as well. Though the nominal structure of the activity was similar, the ways in which students participated was very different.

My interest here is not so much in the best ways to conduct the activity nor in the outcomes in student writing. Those, I think, need a larger data set. Rather, I want to focus on the ways that student engaged with the discussion phase of the activity by contrasting two implementations in which the activity structure was very different, and two implementations in which it was ostensibly the same. The main question of interest, then, will be how to frame a reasonable hypothesis to understand the variance between these three cases. My goal in the end is to use these contrasts as a foundation for creating a strategy for future research that will be aimed at understanding the variance.

## II. REVERSE-ENGINEERING SCIENTIFIC PAPERS – THEORY AND PRACTICE
### A. Theoretical basis of the activity

The standard approach to initiating a "guided discovery" or "scaffolded learning" process is to activate students' prior knowledge (the literature is extensive, but for some of the original work see, e.g., Bransford and Johnson, 1972 and McNamara et. al , 1996). Interactive Lecture Demonstrations (Sokoloff and Thornton, 2004) and ConcepTests (Mazur, 1997), for example, leverage situations in which students have direct personal experience of relevant phenomena but in which the intuitive understandings they have derived from that experience are at variance with physical law (note that "prior knowledge" is a technical term from cognitive science referring to the understandings that people bring with them to a new experience and does not necessarily imply that those understandings are correct).

This approach, however, begs the question "how does one activate prior knowledge in situations where it cannot reasonably be expected to exist?" Physics students have not often had the opportunity to experience phenomena that can be addressed by, say, quantum theory or relativity. Simulations and some experiments can be a partial support but they tend to be difficult to interpret or even to reliably perceive what is happening.

Research-based pedagogical interventions are often described as constructivist as if it were a synonym for hands-on activities. While that may be true, it is not necessarily true. Constructivism is an information processing theory of knowledge formation that describes how perceived information becomes personal knowledge. It does not directly imply any specific assumptions about the source from which that information came. It refers equally to knowledge obtained from a lecture or text as to knowledge obtained from discovery. The critical difference between these two information sources is not so much the nature of the source as it is the existence (or not) of differentiated knowledge structures in the learner that enable the incoming information to be organized into an accessible form. This is as true of information that is told or read as it is of information that is experienced directly. However, when information that is told hits the unprepared mind, in which such an organizing framework does not currently exist, it becomes an end rather than a means, to be memorized rather than used as a tool.

We could speculate, then, that the critical difference between knowledge obtained from direct experience (as in "guided discovery" for example) versus knowledge obtained from being told is that the direct experience not only provides information about the phenomena, it also affords noticing differentiating features of the phenomena that enable that knowledge to be structured into a usable form in memory. But just because phenomenal information and differentiational information *can* be obtained simultaneously does not imply that they *must* be obtained simultaneously any more than the fact that direct experience affords noticing such information necessarily implies that it will in fact be noticed.

The use of contrastive sets as a tool for noticing differentiating features seems to be consistent with this point of view. This idea has a long history in the theory of perceptual learning (see e.g. Gibson and Gibson, 1957; Gibson , 1969). More recently, contrastive sets have been used as an instructional tool for more abstract information. Schwartz and Bransford (1998) expanded the idea in a major way when they used active comparison of contrasting cases as the foundation for instruction in several sophisticated experimental design methods in cognitive science. At around the same time, I was working on physical and Earth science units for Learning by Design (Kolodner et. al., 2003), a project-based approach to middle school science instruction based on complex design problems. The central design problem underlying the Earth science material was tunnel construction and the geologic problems that needed to be solved, something that did not involve much direct interaction with physical artifacts as in the other units. I began working on a method of using contrastive pairs of expert case histories of real tunnel construction in which similar problems were encountered but managed in different ways, along with a simple written tool to help them critically interact with the histories.

The trouble is that while students are capable of noticing distinctions in contrastive sets, they rarely have the overarching mental models necessary to understand the distinctions they have noticed. However, having already noticed distinguishing features, an opportunity may be created to efficiently convey those models to them as learners are now prepared to be told the significance of the distinctions they have noticed. In short, examination of contrastive sets can create "a time for telling" by preparing a place in memory, attached to differentiating structures, and ready to be populated with information. In a clever series of experiments, Schwartz and Bransford were able to show that while performance on recall tasks might be comparable between groups who experienced the contrastive cases and those who did not, the ones with that experience indicated a deeper and more flexible understanding through their ability to perform better on prediction tasks.

It might be objected that the intent of this activity is not very close to that envisioned by Schwartz and Bransford and that is true enough. But their use of contrastive sets did not match very closely that of Gibson, from whom they got the idea. Contrastive sets are a powerful tool with broad applicability.

B. Activity structure

My approach to the scientific paper activity is a variant on the approach of Schwarz and Bransford. There is a similar initiating experience of contrastive reading and reflection to create differentiating memory structures, but the meaning is filled not by lecture but by group discussion and co-construction of a common understanding. The idea is to activate and leverage some of their prior knowledge from non-scientific text, to modify and extend it to the scientific context, and to obtain their buy-in to the results.

The goal of the activity is to help students learn how to structure their textual communications according to the norms of scientific practice (at least in physics). We use the following sequence:

1. Students reach the end of an experimental investigation and are told to write a scientific paper on their results.
2. After handing the paper in, they are given several actual published papers drawn from a variety of sources. Depending on the size of the class, either each group or each individual will receive a different paper. In this activity, group work is of lesser importance compared to the need to analyze a wide variety of contrastive cases.
3. Students are asked to reverse engineer the paper. They are asked to read their paper without worrying much about whether they understand the details of the science, but paying attention to the logic and organization of the paper and how it may be similar to or different from writing they have done in the past.
4. They produce two artifacts from their papers. First, they create an outline, capturing the formal structure, which can usually be done as they read. Then they create a concept map to capture its logical structure.
5. They then have an opportunity to look at and discuss each other's work informally, comparing to their own to notice similarities and differences.
6. We have an extended discussion of what they saw and why the papers might be written the way they are, typically lasting 60 to 90 minutes, and often quite animated.
7. Often at student instigation, we construct the bare bones of a grading "rubric" (really, in practice, more of a design specification). This consists of what elements should or could be present, what characteristics operationally define "good" for each of those elements, and how they wish to be graded.
8. They are offered (and often request – see the quote that opens this paper) the opportunity to revise their papers.

This community-generated rubric seems to result in much more significant changes in performance than does an externally imposed one, even though they may embody similar specifications. Schwarz and Bransford's theoretical description helps to understand why, since the important anchors of the rubric are deeply connected to differentiating features that were constructed by the students themselves.

The activity is designed to afford the opportunity to generate differentiating features that distinguish scientific text from ordinary, narrative text, as well as from their own work immediately prior (step 1 above). While Schwartz and Bransford fitted their features into an overall scheme through a subsequent lecture, we use a discussion. There are a few reasons for this. One is that while students do not typically have much prior knowledge of scientific text, they do have prior knowledge of lab reports and narrative text and these provide important differentiating points which could not be known in advance by a lecturer and which are worth activating and extending. Another reason is that by negotiating a common understanding, and embodying that understanding into a commonly agreed upon rubric, students may become personally invested in conforming to the guidelines they have identified. The importance of this step will become apparent in comparing three recent implementations, but the importance of leveraging prior knowledge and co-constructing shared knowledge has been widely recognized at least as far back as Vygotsky (1978).

### C. Constructing a Contrastive Set of Papers

In order to function productively in this activity, the papers students are exposed to need to be selected with some care. To form a contrastive set they must (a) conform with a common core of desirable target features while (b) exhibiting sufficient variability within those features to afford the opportunity for useful contrasts. We are seeking contrasts in the logical structure and organization of papers, and not so much in the scientific content. Over time, we have evolved a set of papers selected according to these guidelines:

1. The main focus is on experimental physics since students will be writing about laboratory activities. Consequently, we avoid purely theoretical papers.
2. Theory is not, however, totally absent from the set as it plays an important role in both designing interesting experiments as well as understanding the results. We did not wish them to think of papers as simple exercises in data analysis. We therefore paid attention to variations in the balance of theory vs. experiment in the papers selected, and required that, within the domain of experimental physics, the full range of possibilities should be represented, from purely empirical work to work that involves significant interplay between experimental and theoretical analysis.
3. We wish students to learn to construct rigorous, formal arguments so informal papers are not desirable. We had originally thought that articles from *The Physics Teacher* would be useful due to ease of understanding, but we quickly saw that they were in fact impediments due to their lack of formal structure.
4. Distinct and perceptible degrees of expertise should be visible in the papers selected. Ideally, we would have a mix of good and bad papers but, of course, peer review is designed specifically to prevent the latter from appearing at all. We settled on author expertise as a proxy, and drew on the *Journal of Undergraduate Research in Physics*, where peer review is gentler.
5. Papers should be clearly written for a spectrum of audiences. This turned out to be a critical feature in changing student performance.

The common core being highlighted here is the construction of a formal argument based on experimental evidence. Variance is along the dimensions of levels of expertise, specialization for particular communities, and levels of formal theory integration. Table 1 summarizes the important dimensions along which the selected papers vary. The goal was to have at least two papers covering each important dimension. Our collection has expanded somewhat, but these were the papers used in the implementations reported here.

TABLE I: Contrastive Dimensions for Selected Papers

| | | Mauk & Hingley (2005) | Bacon et. al. (2001) | Cheng et. al. (2004) | Coletta & Phillips (2005) | Penner (2001) | Foust (2002) | Easwar & MacIntire (1991) | Cleland (2001) |
|---|---|---|---|---|---|---|---|---|---|
| Audience | Undergrad | × | × | × | | × | | | × |
| Audience | General Expert | × | | × | × | | | | |
| Audience | Specialized Expert | | | | × | | | × | |
| Empirical → Theoretical | Emp. | | | × | × | | × | | |
| Empirical → Theoretical | Mostly Emp. | × | | | | | × | × | × |
| Empirical → Theoretical | Substantial Theory | | × | | | × | | | |
| Author Expertise | Expert | × | × | × | × | × | | × | |
| Author Expertise | Undergrad | | | | | | × | | × |

## III. METHODOLOGY OVERVIEW

The total number of students varies somewhat from year to year but the size of each lab section is fairly stable, as is the student population. Summary data is in tables II and III. Ethnicity was 100% African-American in all implementations, and gender was 100% female except Implementation 1 which was 9% male due to cross registration between Atlanta University Center campuses.

Table II: Demographic data by implementation

| **Implementation** | **N** | **Classification (%)** | | | |
| | | **Freshman** | **Sophomore** | **Junior** | **Senior** |
|---|---|---|---|---|---|
| 1 (Spring 2012) | 11 | 64 | 18 | 18 | 0 |
| 2 (Fall 2011) | 30 (3 sections) | 17 | 56 | 27 | 0 |
| 3 (Fall 2012) | 13 | 0 | 53 | 39 | 8 |

Table III: Distribution of Students by Major (%)

| Major | Implementation 1 | Implementation 2 | Implementation 3 |
|---|---|---|---|
| Physics | 27.3 | 3.3 | 0.0 |
| Physics/Dual Degree* | 0 | 6.7 | 0.0 |
| Math | 9.1 | 13.3 | 30.8 |
| Math/Dual Degree* | 9.1 | 0 | 15.4 |
| Math/Comp. Sci. † | 0.0 | 3.3 | 0.0 |
| Chemistry | 9.1 | 30 | 15.4 |
| Chem./Dual Degree* | 0.0 | 3.3 | 0.0 |
| Biochemistry | 18.1 | 10.0 | 7.7 |
| Biochem./Psychology | 0.0 | 3.3 | 0.0 |
| Biology | 9.1 | 3.3 | 0.0 |
| Comp. Sci | 9.1 | 10.0 | 15.4 |
| Comp. Sci./Dual Degree* | | | 7.7 |
| Dual Degree*/undeclared | 9.1 | 3.3 | 7.7 |
| Undeclared | 0.0 | 10.0 | 0.0 |

* Dual degree is a 3+2 engineering program
† Indicates double major

Implementations One and Three (in-class) were taught by the same instructor, and Implementation Two (online discussion) by a different instructor. The activity occupied a single three hour lab in Implementations One and Three and was distributed in assignments across Weeks 2 to 6 of Implementation Two.

All student-produced artifacts were collected and the in-class discussions were video recorded, transcribed and coded. Coding categories (see Appendix A) were emergent from the data rather than externally imposed expert categories, so they are reflective of what actually occurred rather than a measure of how close the event was to some ideal. The codes are simple counts of specific performances as there is not enough data at the present time to reliably distinguish variations within categories. Coding was done by one faculty member with a check of Implementation One for agreement.

The purpose of coding the video is to characterize two properties: how is the flow of the discussion being controlled (i.e., using the HPL Framework[1] (Bransford et. al. 1999) is it student, community, or instructor-centered), and what are the major resources students draw on to participate in the discussion. Fourteen coding categories emerged that proved to be reliably identifiable. Appendix A contains operationalizations of the codes.

---
[1] For a brief description of this framework, see section six of this paper.

This was not a designed experiment. It was, rather, a set of experiences that in retrospect posed an interesting question. This is why I present it as a case study intended to more precisely frame that question so that a carefully designed experiment can be conducted. At that time, we will have to seriously address the questions of reliability and validity of our coding schemes.

## IV. DATA AND ANALYSIS: A TALE OF THREE IMPLEMENTATIONS

There were three implementations of this activity from which data was collected, two in the form of in-class discussions and one in the form of an on-line forum-based discussion. They collectively tell an interesting story and also raise interesting questions for research going forward. The coding results for the implementations are in Table IV.

TABLE IV. Coding For Each Implementation (1 and 3 are whole class, 2 is online forum)

| Code | Implementation | | | Code | Implementation | | |
| --- | --- | --- | --- | --- | --- | --- | --- |
| | 1 | 2 | 3 | | 1 | 2 | 3 |
| *Discussion Flow Control* | | | | *Collaborative Knowledge Construction* | | | |
| Instructor Prompts | 2 | 6 | 20 | Negotiated Understanding | 12 | 0 | 14 |
| Student Prompts | 12 | 0 | 6 | Intergroup Comparison | 9 | 0 | 13 |
| *Personal Resources* | | | | *Structure Analysis* | | | |
| Accessing Prior Knowledge | 13 | 0 | 8 | Structural Critique | 21 | 0 | 16 |
| Instructor Prior Knowledge | 7 | 0 | 12 | Distinguish Structures | 3 | 0 | 11 |
| Create Example | 3 | 0 | 1 | References Paper | 15 | 0 | 23 |
| Big Picture | 5 | 0 | 1 | Functional Assessment | 1 | 33 | 3 |
| *Affective Response* | | | | | | | |
| Positive Affect | 10 | 13 | 5 | | | | |
| Negative Affect | 4 | 5 | 3 | | | | |

These counts are not quite mutually exclusive. Some properties of the discussion fall into more than one category (for example, big picture discussions are generally a break from the current run of the discussion and so tend to also qualify as student prompts).

*1. Implementation One – Negotiated Understandings in a Conversational Frame*

With 11 student prompts versus only 2 instructor prompts to introduce new discussion topics, the implementation one is clearly community-centered, being driven largely by the students and they are reacting to issues that they raised. They are directing the flow almost all the time and the observations that result emerge from their perceptions of the readings. The role of the instructor is to help them shape and make precise the things they have noticed and the thoughts they have, and to provide some relevant background information on the scientific publication process when that perspective is needed. The instructor is not silent, but he largely reacts to what students articulate rather than prompting them to think about some specific issue. Third, they are negotiating a common set of meanings with one another as can be seen from the large number of intergroup comparisons and negotiated understandings, and they are actively trying to integrate this knowledge into their existing understanding both by drawing on elements of prior knowledge to apply here as well as trying to fit what they are learning in their overall perception of their goals, both at the course level and the career level.

I would like to focus particularly on the role that student prior knowledge plays in the negotiation of a common understanding since this is one place we can see most clearly some of the existing resources that students are drawing on to understand what they are seeing in the structure of these papers. In particular, there are two points in the discussion where students reach well beyond the boundaries of their physics course to import some ideas from literary analysis in their English courses. How they do this, and particularly their apparent motivation, I think tells us something important about how to think about their reasoning process.

Approximately the first fifteen minutes are informal discussions between groups as they compare their outlines and concept maps. The instructor calls them together with an open-ended prompt (I is instructor, S is student with numbers distinguishing different students):

> I: Tell me what you noticed – similarities, differences, comparisons with what you've done in the past?

There followed about two minutes of various students raising small scale issues, and then one student kicked off a major discussion episode.

> S1: I have a question for the group. The biggest difference I saw regarding the outlines, we all did basically the same things, but some lumped things together. What is better? Should we have it all together and make it flow, or have it separate and clear?
>
> S2: I vote for flow. You can't make it flow and sound crazy. If you chop it up, it is hard to make sense.
>
> S3: I'm not a physics person, but in chemistry we chop it up. . . Intro, theory, materials and methods . . .

Though it was articulated a little vaguely, they immediately picked up on the central issue – should we tell the story of the experiment, or should we take all items of a similar sort and group them into the same section? It isn't entirely clear what motivated the question, but it quickly led one student to recall the practices in her major. They continue to elaborate this contrast between physics and chemistry.

> S3: Comparing physics to chemistry – this is just how I feel – physics has more real world concepts that it relates back to whereas chemistry, it's more focused so that's why it's chopped up.

> S4: I'd just like to say chemistry is very real life [laughter].
>
> S3: Physics is more concepts. Chemistry is more . . . like . . . detailed.
>
> S4: I get what you're saying. Physics is like trying to figure stuff out, in chemistry it's more like proving what we already know, doing what we already know.

Leaving aside the interesting perspective on their chemistry courses, they are clearly trying to articulate a perception that physics has a stronger conceptual motivation, which might account for the disciplinary differences in the construction of papers. The instructor picks up on this viewpoint.

> I: . . . Physics is interested in fundamental laws so the inclination is to set up simple experiments so you can see the laws clearly. Often, in chemistry and biology, you're dealing with what the world handed to you, and there's a lot less ability to manipulate . . . There are differences from one discipline to another and how you would approach.
>
> S1: Question: in our paper they pretty much spaced out graphs, formulations, calculations, throughout the whole paper where I would normally group that together in a calculations place [sounds of agreement].
>
> S5: Our paper had one question, three subquestions, and every time he would touch on something that he had evidence of then he would put it right there next to it. . . I think it's best. He had like three different experiments so if he put it all on one page, how would we know the significance of each?

S1 is still wrestling with the motivation for narrative flow in the physics papers versus thematic grouping in her prior experience. There is some back and forth, comparing the approaches in their papers on the placement of various items, resulting in a negotiated compromise view to place figures in the narrative flow with labels for reference as needed at other points in the paper. The instructor again draws an analogy from his own experience:

> I: I frequently work with biologists and for someone with a physics background it is sometimes like reading papers that dropped down from Mars, the exact same structure with the exact same headings in nearly every paper you look at. This made it very confusing for me because understanding the data involved looking at some of the analysis but that didn't come until much later so it would refer forward and backward.
>
> S2: Oh, so not in order, you have to jump around a lot. I wouldn't do that, I think it should be more in order . . .
>
> I: Why do you think someone would do it differently?
>
> S4: Culture, (another S simultaneously: more convenient for them) it's got its own culture, and as you've seen we distinguished between physics and chemistry. Maybe if that's the way that a lot of the literature is then you want to concede to that pattern. Like writing in an English class vs. a science class, that's going to be a totally different style.

This is the first instance of a fairly distant reference to English courses and leads to importation of a major set of ideas from literary analysis, the first being the impact of culture on forms of expression, and the assumption that different scientific disciplines have somewhat different cultures. This comment led to another student expressing her frustration over understanding her paper (a short but dense one on special relativity). She prefers a completely self contained explanation in every paper:

> S3: I think it's important if you produce a paper anybody should be able to understand what is going on even if you aren't in the discipline. That's the issue I have with scientific papers, I'm not the, like, a reader/writer kind of person, I need things to be clean cut, so even someone who's not in college could sort of understand instead of throwing around these big words, big numbers, without explaining.
>
> I: Was your paper like that?
>
> S3: I think they did explain some things, but not my level of understanding, I need some pre understanding of what this is.
>
> . . .
>
> I: When I was a graduate student, just starting work on my research, I used to complain about this to my advisor. It seemed like every paper, you couldn't understand it until you read the papers it referred to and you had to go back through 20 years of papers to understand anything.
>
> S3: I think this needs to be explained like this is your only source.
>
> I: So what did the other papers look like, did they assume you had some knowledge you didn't have or were they self contained?
>
> S2: It assumed you had some prior knowledge. It has, like, sigma and doesn't say what it is.
>
> S4: Mine did a little bit. Mine was about earthquakes. I come from a place where earthquakes are common. My house is right on the San Andreas. When we go to school, we learn everything about earthquakes. So I think it depends on your knowledge. Someone from where there aren't many earthquakes might be a little more confused.

Here, both the instructor and one student have referred to an idiosyncratic bit of prior knowledge which, in this instance, serves the usual turn-taking conversational function of extending the discussion through an analogy with personal experience.

> I: Why do you think they're not self contained?
>
> S4: Probably for a particular audience
>
> I: That's true
>
> S1: I had an English teacher wanted me to tell what the story is about to explain your ideas, but another said I've got the book, I know what you're

> talking about, just give me the facts, your opinions on it. I think this is one of those things where if you're writing it for a professor, he's like I know the background, just give me the facts.

Here, the idea of writing for a specific audience is imported from English classes by two different students. This turned out to be a major organizing idea that was returned to multiple times over the course of the remaining discussion.

> I: So who's the audience for your papers?
>
> S3: I think mine was like a board of physics people, like people over a physics department in a school.
>
> . . . .
>
> S5: in this person's defense, he put in like a keyword thing. "You're supposed to look these words up before you read my paper."
>
> I: That's really a way of doing a search for related papers. But that's an interesting way to think about it, though. It's sort of an encapsulated description of who the audience is, right? If you understand these, you're part of the audience.

Several very important points about writing papers have emerged organically from this discussion. Students have noticed first that a paper is embedded in a culture, and that the norms of that culture can dictate forms of expression. They draw on their experiences from analyzing narrative text in English classes to adapt this idea as well as the idea of writing for an audience, sharing a common background knowledge, which accounts for why papers are not self contained as well as makes them difficult for outsiders to comprehend. Much of this was instigated by S3, who was initially quite frustrated by her paper, on a relativistic time dilation experiment, which she found difficult to understand. She wanted papers to be self contained and it was her grappling with her frustration that pushed the group toward conceiving of writing for a well defined audience. In the end, her experience pushed the group to negotiate a common understanding of why they should write for an audience and who their audience should be.

I quoted extensively from this part of the conversation because it illustrates how a negotiated understanding was arrived at, as well as their fluid use of prior knowledge from personal to disciplinary to outside the sciences altogether. This was a persistent feature of the remaining discussion, with most prior knowledge references being disciplinary.

The literary notion of writing a *scientific* paper (as opposed to an English assignment) for a particular audience was a novel one for these students and it appeared again over an hour later. The context was constructing a grading rubric and trying to specify measures of quality for the important pieces of their papers. They began with discussing what makes a good title. Most of them had written drafts with titles like "Lab 1" or "From the middle to the end!" which, if they have an idea of audience at all pretty clearly are aimed at the instructor. There are some ten minutes or so of discussion revolving around summarizing the argument, stating a question, or the main conclusion, the main point being, as one student put it, "to give insight into your argument." At this point, a student reaches once again for prior knowledge from English courses:

> I think like in English we used to always read peoples, people, like, writing, and she would say like "analyze their rhetorical devices" and how it seemed beneficial to the audience, and so, like, a good title of that paper would be "the benefits of such and so's use of rhetorical devices in their work" and if you think it is a benefit, like your argument is that you think it benefitted their argument then you would say that but if your argument was it was a downfall to their argument, then you could say "the setbacks of their using rhetorical devices, " stuff like that.

Here, they have changed their viewpoint from reporting results to the instructor to professionally communicating arguments to other interested parties. They've begun to develop an understanding of a potentially broader audience and conceive of a title as a means of accessing that audience. Again, this conception was imported from literary analysis in English courses and there was no evidence in their class work for such an understanding prior to this discussion.
I will return to these excerpts in section V below, but first I want to discuss two other implementations.

## V. IMPACT OF CHANGING THE FRAME – IMPLEMENTATIONS TWO AND THREE

*1. Implementation Two – Online discussion forum*

Implementing this activity as designed occupies essentially an entire three hour lab period. We therefore made an attempt to streamline the course a bit by moving the activity outside of class, taking advantage of an on line discussion forum feature in our institution's course management software. There were several salient changes in the nature of the activity.

First, it began at a much earlier point in the course, in week 2 rather than approximately week 6. This was well before they were ready to write a paper, so the direct connection between the activity and what they were currently doing was significantly weaker. Second, prompts were distributed as items on assignments that also contained more traditional data analysis and planning questions. Third, the nature of the prompts changed to be somewhat more directive. Contrast, for example, the initiating prompts. In implementation one, it was conversational and open ended:

> I: Tell me what you noticed – similarities, differences, comparisons with what you've done in the past?

This triggered a vigorous interactive discussion that led directly to the issue of cultural differences between sciences within the first 5 minutes, without further prompting from the instructor. Throughout the remainder of the activity, the instructor deliberately tried to follow the discussion rather than lead it, as can be seen in the number of instructor versus student prompts. In contrast, the initiating prompt for the Implementation Two was much more explicit on what to think and how to think it:

- Begin to learn about writing a scientific paper.  Read the scientific papers posted on WebCT.  You may not fully understand them, but you are reading to see how you should write a scientific paper, not for understanding the paper fully.  Do the following:
    - List the general common features of a scientific paper.

- Examine one of the papers to determine how the paper supports its conclusions. Create a diagram showing, for each conclusion, how it is supported by analysis and how, in turn, the analysis is supported by data and results. Also determine how each conclusion relates to the introduction in the paper. This sort of looks at the paper in reverse.
- Create a rubric for grading a scientific paper for this class.

Students were told on this and subsequent assignments to post their work in the online forum and respond to one another's observations, revising their diagrams, rubrics, etc. as needed. In addition, they were reminded by email at intermediate points between class meetings to return to the forum. They were encouraged to react to and elaborate on the discussion. On a later assignment, they were provided with the rubric that would be used for grading and asked to compare it to their rubrics and use it to assess the paper they had read. The only point in Implementation One where the instructor approached this level of explicitness in his prompting was late in the discussion when he directed their attention to the way that data is presented (raw data or means and standard deviations?), as they had not talked about that yet.

Students responded to the letter of the assignments and little more. In particular, there were no observed instances of interactions between students despite multiple prompts to do so. Their responses tended to be affective, for example when they were provided with a rubric and asked to assess its suitability. The intent was that they would compare the rubric to their own and to the structural features of the papers they had read. Instead, they responded:

> S1: The rubric for writing a scientific paper was the most clear rubric I've seen in my opinion. It was very clear in that it specifically broke down all the components that a good scientific paper should have. I feel it was clear because under the "4" column it basically tells you what must be included in order to get full credit.

> S2: The rubric is similar to a guide of what we need to make sure that we include in our scientific paper. The fact that things are weighted differently, you have to make sure that you put forth more effort or more time on the things that take up a bigger percentage.

Applying the same coding scheme as above to the online responses is instructive. The results are in Table IV above.

The difference implementations one and two is striking, and even more so if one realizes that the 33 functional assessment responses come entirely from the single assignment on which they were told to apply the provided rubric to their papers, and they did so. Every spontaneous student response was affective, mostly positive.

The asynchronous nature of the online discussion is likely an issue. It has been known for over a decade (see Brown and Duguid, 2000) that interactive technologies tend to be more effective at maintaining existing communities than in creating new ones. This activity is intended to be community centered but in Implementation Two it appeared at the beginning of the term, before a community and its norms could reasonably be expected to exist, in contrast to Implementation One where it appeared six weeks in. But I hypothesize that the change in the nature and means of distribution of the prompts was at least as important a factor, and triggered a framing change.

A frame is a set of expectations for an event based on a history of similar events (see e.g Hammer et. al., 2005). So what is the basis on which "similarity" might have been judged in this case? The change in the nature and distribution of prompts is an obvious difference between the two implementations, as is the greater social distance from their peers imposed by the online environment. By changing from open-ended to task-based prompts, and by distributing them along with more traditional laboratory problems as part of weekly assignments, it is likely that we inadvertently caused students to frame this activity as a homework assignment rather than a conversation. They then responded to the instructor and not to one another.

The responses to the rubric prompt from Implementation Two, quoted above, are a good example. They take the information provided by the instructor as a given, not as a starting point, and proceed to state its positive attributes. There is no interaction whatsoever between students, and none of them ever responded to any other students' comments, despite being prompted to do so. The codes in Table IV indicate that the entire activity proceeded in similar fashion and was primarily, in the HPL framework, assessment-centered.

In implementation two, there are responses to questions on an assignment with no evidence of interaction between students. In the implementation one, there is turn- taking, recalling analogous experiences, negotiating modifications of understanding until they arrive at a common agreement. These clearly embody radically different perceptions of the situation these two groups perceive themselves to be in.

*2. Implementation Three – In-class discussion*

In the year following these two implementations, there was a third that was also in a class discussion format. Its results lie in between the previous two, as shown in Table IV. The activity structure was ostensibly the same but the course it followed was not. So we will have to look beyond the activity itself to understand this variance.

What stands out here, compared to Implementation One, is the inverted ratio of instructor to student prompts. The discussion flow is much more controlled by the instructor (the same one, as it happens). What is not present in this coding, however, is a subtle difference in the style of discussion. Again, an excerpt from the discussion will provide the needed perspective. The issue of audience arose here as well, and the dialog surrounding that is representative. It is important to note here that S1 – S4 were all members of the same group. S5 and S6 were members of a different group. Once again, it starts with some frustration:

> I: So what else? Anybody see anything unusual, confusing?
> 
> . . .
> 
> S2: I think mine, it had way too many, like, other experiments. Like they had way too many like they had 10 other experiments in their paper
> 
> I: That they did or that they referred to?
> 
> S2: No no, no, that they referred to that somebody else did. And I felt like, they introduced it in the beginning and I really liked it but then when I started going through the paper it was like they kept bringing in more and more and it was like, ok, I get what you're trying to say, you don't really need to keep, you know, using different experiments to prove your point. And I thought it like made their argument less meaningful because it was like the whole paper was comparing different, um, . . .
> 
> I: Why do you think they did that?

S2: Um, well, basically, They were trying to prove why those experiments were wrong but they all gave the same reasons for why they were wrong so I felt like they didn't need that many. Some of them might have been different reasons, but I felt like, if, if you include an experiment and you say why it's wrong you don't need like 5 experiments just to say that they all were wrong because x, y and z. Like the same x, y and z.

With some continued probing, the issue of audience arises but, in contrast to Implementation One, it isn't clear why.

I: So why do you think they would spend so much time talking about what everybody else did, going into that kind of detail on it?

S3: Maybe trying to compare?

I: OK so they are trying to compare, but who benefits from that?

S1: They do.

I: How do they benefit?

S1: If they're trying to prove something that the other experiments, like if they're trying to prove the opposite of what the other experiments proved, I guess, and show that they're false and that like they're really just showing that what they're trying to prove is true. So the more that they prove it's false they're just making their argument even stronger.

I: Ok, so, who are they proving that to?

S2: The audience.

I: What audience?

S1 and S2: The reader.

I: who's the reader?

S3: Me.

I: You? Well, I guess you are by definition right now, but are you the reader that they probably had in mind?

S1 and S2: No.

I: Who did they have in mind?

S2: Probably the person, mine was about, like, how a flute, basically, different materials could change the tone, so probably, like (simultaneous with S4: Music industry) musicians.

I: Musicians, or maybe people who are interested in acoustics? Engineers?

S2 then draws on prior experience, but largely to explain her frustration.

S2: Uhhhh . . .I think . . . I don't know. I guess maybe this is kind of hard for me because I played the flute, maybe, so (laughs) that might count, but I feel like even if I didn't I feel like anybody that read this article could of, like, gotten the gist of it after a couple of experiments.

. . .

I: Ok, well, I'll tell you that not all of these papers are good papers.
S4: yeah some were bad.

I: It's really pretty difficult to find bad papers that have been published, but at least a couple of them were written by undergraduates . . .

S2: this one seemed like it was written by a middle schooler (general laughter) You know, really, like, I was like, it was so bad to me, I just didn't understand why, it was like every different paragraph was like "In 1920, in 1930, in 1940 . . " and it was like . . .

I: Yeah, ok, but that's a fair critique, and the person that wrote that was an undergraduate, so you can assume this was all new to them. So ok so there's an audience for the paper but the author doesn't always hit that audience exactly, maybe they should pay more attention to it . . . so did that play a role in any of these other things, that you noticed? Some of them you said were hard to understand, what made them hard to understand?

The discussion wandered away from the question of audience for just over three minutes before coming back.

I: anything else that seemed to be out of place, or missing?

S5: I think in my paper, which was on relativistic time dilation they didn't really connect the relevance between the experiment and the actual topic.

I: so what was the experiment?

S5: So it pretty much was based upon the count of muons but they didn't really say what time dilation has to do with the count.

I: OK, did they discuss time dilation at all?

S5: Barely, but they had a minor, like a paragraph but it just gave a general definition, but in the conclusion they did say that the concept of time dilation is difficult for a student in the first modern physics course, so I guess they pretty much said you won't understand it if you're taking that course, or not taking an advanced class.

I: Ok so it sounds like they were writing about . . . first of all, who was their audience? Who were they writing to?

S5: Yeah (pleased expression and nodding head) probably students in the advanced physics class.

I: Are they writing to the students or the instructors?

S5 and S2: (simultaneous) Instructors.

I: Sounds like they were writing to the instructors. So what could they assume about the knowledge that their audience has?

S6: Better know it if they're gonna teach . . .

These excerpts illustrate an important point – in Implementation Three, interactions were almost always between the instructor and a specific group. There was little unprompted cross-group interaction, and most of the time, even within a group, the discussion episodes were largely between the instructor and a specific student. Indeed, other groups were as likely to be working on other tasks or even taking restroom breaks as they were to be listening to, let alone participating in, the discussion. We could fairly describe the discussions as conversational (Implementation One), assignment (Implementation Two) and Socratic (Implementation Three). Whereas the difference between One and Two may be fairly attributed to the radical change in prompting, the difference between One and Three cannot. The instructor is engaging in a similar way but the students are not.

I'd like now to propose a way of understanding the differences in the way they engaged in the activity.

## VI. DISCUSSION – A TALE OF THREE MODELS

I have been using the language of resources and framing in this paper, for obvious reasons. Resources/framing (R/F hereafter; Hammer, et. al., 2005) is a quite popular viewpoint in physics education research but I would like to take a moment to describe two alternative models of learning, Case-based Reasoning (CBR) and the ecological perspective, that may help both to understand these observations and to frame a research agenda going forward. I tell my students the more ways you can think about a problem, the more deeply you understand it. The same is true for me. My goal in this section is to establish the foundation for a research strategy to address the question "How do these students choose a frame?"

Case-based reasoning (Kolodner, 1993 and López de Mántaras, et. al. 2005 for overviews) has roots in intelligent systems and specifically the attempt to construct accurate computational models of human learning and reasoning processes. It can be thought of as a model of reasoning by analogy with prior experience, a sort of experiential descent of ideas with modification. CBR is founded on two assumptions about the world of learning, that similar problems tend to have similar solutions, and that the problems we encounter from day to day tend to recur so it is worth remembering and adapting past solutions. Indeed, it is supported by a vast body of evidence indicating the importance of remindings in human reasoning. CBR includes detailed descriptions of mechanisms for injecting creativity to adapt previous solutions to variant circumstances. In a sense, it serves as a model of inductive learning, but in a way that takes seriously the importance of prior knowledge and the role of failure as well as success.

CBR divides more or less naturally into two classes:
- Interpretive CBR: uses prior cases as reference points for characterizing new situations. In the R/F model, this would be akin to a process for activating a frame.

- Problem-solving CBR: uses prior cases to suggest solutions to novel circumstances

Case-based learning can be thought of as an iterative cycle: given a problem or task to be solved, extract a characterization to use as a probe into memory to retrieve similar prior experiences; adapt those experiences as needed to suit the requirements of the current task; apply that candidate solution and obtain feedback on successes and failures; revise accordingly; repeat as needed and save the outcome as a new case. A central problem, then, is describing how to index experiences so that they can be successfully retrieved as appropriate, and this is a subject to which much thought has been given.

The orientation of the R/F model is situational – a frame is a set of expectations about the situation in which one finds oneself which then triggers the activation of a set of resources appropriate to that situation. The orientation of CBR, on the other hand, is primarily toward the task or goal that leads to the activation of resources that have proven successful in similar tasks in the past. These are clearly related points of view. Goals can obviously be incorporated as aspects of a frame, and situational cues can be part of the information processing that enables case retrieval. But there is nevertheless a clear difference in emphasis in the two models of reasoning, allowing each to contribute differently to a more complete explanation of reasoning phenomena. There is one salient fact regarding the ways that students used their prior knowledge in Implementation One. Each time they reached far afield to import ideas from literary analysis into the dialog, they began with a task orientation, which is characteristic of Case-based Reasoning. In the first occurrence, it was a question raised by a student about whether they should sequence their paper with a narrative flow or in thematically related sections. In the second instance, it was an attempt to specify how to create a title that speaks to a defined target audience.

While the R/F point of view accommodates fluid shifting of frames as an individual tries to fit a current situation to experience, I have difficulty using it to account for cross-contextual learning and especially cross-contextual remindings. CBR, which came out of an attempt to understand how humans comprehend stories and views memory as a hierarchical organization in which lower level components can be shared, helps me better understand cross-contextual learning and reminding. Cases, in CBR, are interpretations of experiences and included in them can be not only solutions to problems but also a trace of how those solutions were derived. CBR's foundational theory suggests that cases are stored in memory under the structures used to process them, inextricably linking and binding together processing structures, generalized memory structures, and episodic memory (Leake, 1996). A byproduct of this linking is that one's experiences, or cases in memory, become accessible when the processing structures used to process them in the past are again activated in current situations. From this, a mechanism can be derived for how spontaneous remindings arise: they arise because a new situation similar to a previous one is processed by some of the same memory structures, and the special characteristics of the new situation match, in a way critical for processing, characteristics of the episode one is reminded of. That is, indeed, the reason one is reminded.

Because I am to look for evidence of framing primarily in discourse (Hammer et. al, 2005), the R/F approach serves me well as a means of thinking about past events. However, its focus on the locality of resource activations (both to the individual as well as the situation) makes it difficult for me to predict outcomes or understand general reasoning processes. This is the purpose for which I find CBR useful. Due to its roots in intelligent systems and the need to program computational models to reproduce human reasoning, CBR must necessarily be a great deal more precise and explicit about the processes by which memories are stored, accessed and adapted, and that enhances its predictive value. I readily admit that these are operator-dependent statements, depending on my own peculiar set of cases and frames, but it indicates how I view them as complementary points of view.

In the present situation, CBR suggests that we begin thinking of the paper writing activity in terms of the overall goal of understanding a text. Activating that processing structure naturally leads to the activation of prior situations that shared that goal, for instance in literature, public speaking or history (all of which appeared in Implementation One and may be implicit in Implementation Three). It is therefore natural to activate the resources that proved useful in those contexts and adapt them to the current situation, therefore leading to the speculations about audience, culture and so forth. Implementation Two recast the goal as providing the answer to a homework problem to one's superior, which activated a largely different set of resources, and the notion of audience in particular never arose. Had I been a more competent operator of CBR, I might have predicted that and avoided this implementation altogether.

R/F, on the other hand, suggests that we begin thinking about Implementation One in terms of a conversational frame but while that naturally leads to the turn taking, analogizing, and negotiations of understanding that accompany all conversations (and which I observed quite a lot of in the coding), it does not naturally help me to understand the importation of ideas from frames that have nothing to do with physics. I could narrow the frame definition to conversations regarding text specifically but that in turn illustrates my problem with operating R/F – I look primarily backward on what happened, using the past event itself to construct powerful and flexible phenomenological descriptions, rather than forward on what will happen, and how the underlying reasoning processes will cause it to occur. Implementation Two made an attempt to provoke the same or similar social aspects of a frame using online discussion forums. Clearly, this failed but I'm not sure I would have been able to predict that no matter what my level of R/F competence is. While R/F does shed some light on the differences between Implementations One and Three ("I'm talking with my peers" versus "I'm answering my instructor's questions"), it does little to help me to understand *why* there is this difference. But, then, neither does CBR.

Ecological psychology is a title claimed by at least two distinct but related schools of thought, following Gibson's work on perception (1979, 1983[2]) and Barker's (1968; Schoggen, 1989) approach to behavior. Both emphasize the role of the external environment in shaping behavior. Gibson's perspective has been described as "ask not what's inside your head but what your head's inside of" (Mace, 1977) and was a radical divergence from the emerging constructivist and cognitivist schools of the time in that he rejected the notion that we process meaningless physical sensations to create meaningful internal representations and instead argued that perception *is* information, namely, the affordances the external environment offers for action. Affordances capture the mutual relationship between the sociophysical environment and the intentions of the actor. More recent developments in embodied cognition (e.g. Gibbs, 2005) and enactivism (Varela, Thompson and Rosch, 1991) may be seen as supporting Gibson's basic position. As Gibbs puts it:

> Our sense of ourselves as persons that endure through physical … and mental … changes is primarily base on our bodily interactions with the physical/cultural world. .. our perception of the sensory world is given to us directly by "affordances." An affordance is a resource that the environment offers an animal, such as surfaces that provide support, objects that can be manipulated, and substances that can be eaten, each of which is a property specified as stimulus information in *animal-environment interactions* [emphasis added]. Each person/animal has a vast set of possibilities for action, based on the perception of affordances . . . Oneself and one's body exist along with the environment, they are co-perceived. (Gibbs 2005)

---

[2] Physicists would find Gibson's analysis of the interplay between geometric optics and perception fascinating.

From this point of view, then, ecological psychology (and, by extension, embodied cognition) is not, or at least not fully, a cognitivist, constructivist model. Affordances span the gap between individual and environment. They have a foot in both worlds, existing partly in the perceptual, sensorimotor system and partly in the changing sociophysical environment.

In *How People Learn* (Bransford, Brown and Cocking, 1999), Bransford advocated what he has since (Bransford, Vye and Bateman, 2000) called the HPL framework for analyzing a learning situation. It, along with a way I propose to extend it, is illustrated in Figure 1.

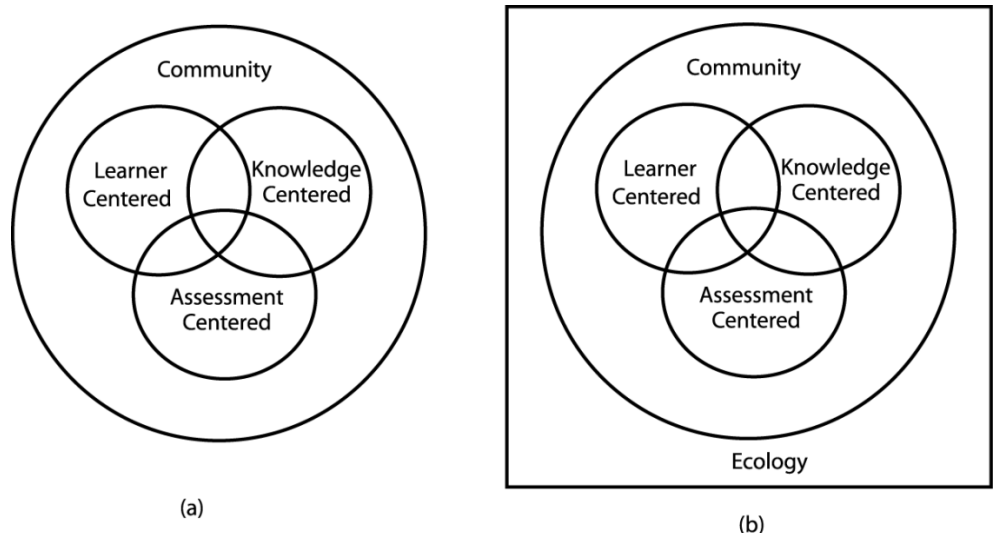

*Figure 1.* The HPL Framework (a) original version and (b) modified version

The original framework covers the interactions between learners, instructors, activity structures, etc. It embodies a cognitivist, constructivist perspective (as, indeed, do R/F and CBR), that the appropriate level of analysis is the knowledge created in the learner's head. This extension embodies the ecological perspective, that the appropriate level of analysis is the relationship between the individual's behavior and the external environment, that knowledge spans both the learner's mind and its environment, that meaning resides in the interaction (this is the reason for the added emphasis in the quote above). And the surrounding environment influences behavior in sometimes surprising ways. There is, for example, a large body of odd results in psychology that come under the general heading of nonverbal communication,[3] results that don't have a neat home in any recognized subdiscipline. In one of these experiments (Ward 1968), participants were seated asymmetrically around a circular table, with two seats on one half and all the rest on the other half. Those on the low density side were visually central in that they received more undivided gaze, and when participants were seated in the visually central chairs, their behavior changed. They talked more, they assumed more responsibility, group leaders tended to emerge from that side. One could equally well say if you are seated on the less visually central side, your behavior changes in that you cede leadership. It is worth noting that in Implementation One, the instructor was seated among the students and moved back and forth to the board as needed to record ideas. In Implementation Three, he stood in front of the class, at the board, mostly continuously. It may well be that physical arrangement of participants and visual prominence of some of them plays an important role in adopting a frame and thereby shaping behavior. However, that is not a factor included in the HPL framework since physical placement in space is not part of either the learner, the knowledge or the assessment.

---

[3] I have a paper on this that has been hanging around for two years since I can't figure out where to send it.

This analysis of the way the individual interacts with their external environment is where Gibson's concept of affordances comes in – understanding this interaction leads to predictability. The environment of Implementations One and Three is ostensibly the same but in fact it is not. The instructor is as much a part of the students' ecologies as they are of his, and what they each do is in response to what the other does. The event is a dynamic relationship, not a static environment. The discourse level analysis advocated by R/F may reflect these differences but doesn't provide the cause. Is it something the students did, something the instructor did, or something the interaction did?

This activity was designed to be community-centered. Implementations One and Three are but there is nevertheless a clear difference between them that the cognitivist, HPL framework does not accommodate. From the extended HPL perspective, neither the knowledge nor the assessments change so the cause of variability is likely to lie in the learners or their interactions with the surrounding ecology or both.

This affords the creation of an action plan for future work by using CBR and the ecological perspective to reformulate the main question "What causes the frame variability?"

- CBR prompts me to investigate the extent to which the current experience reminds them of previous experiences of a similar sort as that will be a measure of the degree of overlap in memory indices. But "of a similar sort" covers a space that includes nonscience experiences.
- The ecological perspective prompts me to investigate the relationship between behavior (both instructor and students) and their respective external sociophysical environments.

The first question suggests post-session clinical interviews to determine what prior experiences they were reminded of, perhaps coupled with a rating scale to measure the perceived relative importance of those reminders. For instance, discourse analysis will not tell us why "audience" arose in Implementation Three as the student who raised it was not forthcoming about her reasoning. The second point is a little more difficult to get a handle on but some headway might be made by revisiting an older ecological concept, Barker's (1968) notion of behavior settings. This does not seem to have been an active independent area of research since the mid-1990's but has been subsumed into many other subdisciplines of psychology as the importance of context has been more widely recognized.

A behavior setting is a sociophysical ecological unit, internally interdependent and distinct from adjacent units, and containing one or more standing patterns of behavior. Affordances couple the properties of the ecological unit to the behavior. Barker identified eleven key attributes of a behavior setting which serve as a credible starting point for coding. Not all of them are important for a strictly classroom context and the list seems to be missing a few things that are at least potentially important. As a starting point, I propose coding:

- Penetration: the degree to which each individual is involved in the setting, ranging from onlooker to leader, and providing both individual and average depths of penetration.
- Action patterns: the functional attributes of patterns of behavior – portions of the coding scheme developed above serve here.
- Occupancy time: at individual and group levels, the fraction of time engaged in the activity
- Behavior map: a map of the location of each individual coupled to their behavior description.

- Discussion frame: measured along the axis individual/assignment to dyad/Socratic to conversational/community. Again, part of the coding scheme here presented seems to capture this reasonably well.

Behavior setting analysis can be done continuously but it is much more common to follow a sampling strategy, coding, say, one minute segments at regular intervals. That is due to the fact that behaviors are relatively stable if the setting is stable so there isn't much to be gained from continuous coding.

We are steadily collecting more examples of implementations, and they tend to lie on the frame spectrum identified here from assignment to Socratic to conversational. Since these seem to be degrees of "community-ness," that suggests the use of a rating scale to generate an overall characterization of each event.

## VII. CONCLUSION

While I still want to collect more data before drawing any final conclusions, early indications are that student performance in writing their papers parallels the character of the discussion along the assignment/Socratic/conversational axis. In retrospect, the attempt to make more efficient use of class time can be viewed as a version of a much more extensive experiment in streamlining conducted by Clark and Linn (2003). Their conclusion was that decreasing instructional time by decreasing the time spent on reflective activities (while maintaining time spent on investigative activities) significantly diminishes the knowledge integration goals they were targeting. We can now extend this. While Implementation Two probably did significantly decrease time on reflective tasks, it was not designed to do so. Rather, reflection is a natural part of conversational interaction through turn taking, comparison with personal experiences, and the like. Interaction increases the likelihood of reflection. By provoking our students to frame the activity as an assignment, we apparently caused them to engage with it in a very different way, as a personal response to a superior rather than an interactive construction with their peers. We did not subtract any reflective opportunities or goals for reflection. So it seems inevitable that it was the enveloping social context that must have made that difference.

Clark and Linn also did not look at the variance within a particular design. While Implementation Three had a greater degree of interaction, it was mostly bidirectional rather than multi directional and the flavor of the interactions was distinctly different. This conclusion goes well beyond that of Clark and Linn, where the reflections were always in response to prompts. Implementation Three is similar to the Clark and Linn experiment but that raises yet another question – the formal structure of the activity was unchanged between Implementations One and Three but the way students engaged with the activity is quite different and we need to account for that variance. If the reason for that variance is not to be found in variations of the activity or the assessment, then it must arise out of the interaction between instructor, students, and their surrounding environment.

A framing difference is at least a compelling hypothesis and seems a natural interpretation of these case histories, but that begs an important question – how is a frame chosen? The Resources model largely depends on discourse analysis after the event (Hammer et. al. 2005) to determine how a frame was chosen. But Case-based Reasoning and the ecological perspective provide some plausible explanatory mechanisms for understanding how a frame will be chosen, as well as some systematic ideas about how to influence that choice. A multiple models approach is therefore more illuminating than any single one. It suggests a strategy of fine grained analysis of a complex system to explore the variability in how instructors and students engage with the opportunities their environments afford.

This investigation has raised more questions than it answers. It is clearly possible to create well differentiated prior knowledge resources regarding forms of scientific communication where none or few previously existed, and to do so using materials (such as text) that have traditionally been viewed in physics education research as of lesser importance than direct engagement. The key idea is to generate a need to know information which will arrive later, in this case prepared through the use of contrastive sets to highlight essential features and filled by a discussion to understand those features. One wonders if this strategy could be used productively in courses such as quantum mechanics, where the phenomena themselves are not readily accessible or where the means of accessing them is so complex as to interfere with reflection on the experience. On the other hand, it is also obvious that the contrastive activity is not guaranteed to produce the desired results, that there are other influences on outcomes that need to be better understood.

## ACKNOWLEDGEMENTS

I would like to thank Keron Subero for allowing me to take over his class for a day, helping to select useful papers, and productive discussion of the results. I would also like to thank the students of the Spelman College Physics 151 course, Spring of 2012, for being so clever and engaging, as well as the members of the Solo PER community for serving as a virtual research group and providing feedback. Finally, my gratitude to Janet Kolodner for assisting with the discussion of case-based reasoning, and Jacqueline Gray for help with the ecological perspective.

---

# APPENDIX A: CODING SCHEME

| Code | Operationalization | Example |
|---|---|---|
| Accessing Prior Knowledge | Uses specific knowledge (correct or incorrect) from outside this discussion to make or understand or illustrate a specific point | I had an English teacher who wanted me to tell what the story is about to explain your ideas, but another said I've got the book, I know what you're talking about, just give me the facts, your opinions on it. I think this is one of those things where if you're writing it for a professor, he's like I know the background, just give me the facts. |
| Intergroup Comparison | Without prompting compare papers or analyses of papers | I have a question for the group. The biggest difference I saw in the outlines, they all did basically the same things but some lumped things together. What is better? Have it all together and make it flow or have separate and clear? |
| Structural Critique | Analysis of the meaning or role of a structural element in either their paper or papers in general | I think the title should kind of mimic the thesis, kind of, not say it verbatim but give insight into what your argument is |
| References Paper | Specific reference to their paper to make a point | Um, do you all have theory sections? Ok, cause we have a section called "theory" so that would make me think that if you're going to reference on this it should be theory and background and all that you did. |
| Big Picture | Explicit break from the general discussion to take a more global perspective | I'm confused, I guess, between the different sections. Because I feel like in lecture we're supposed to show what we're doing step by step, every calculation, that's to check to make sure our formulas and our math and such is right. And you know, the structure of our papers are a little bit different. How are we supposed to relate what we're learning in lab and this structure and style and important key points to lecture? |
| Instructor Prompts | Statement by the instructor that marks a specific and identifiable break from one topic of discussion to another | OK, so you guys have done a lot of lab reports in your life. Did you notice any similarities or differences? (Immediately follows discussion of whether engineers write similar kinds of papers.) |
| Student Prompts | Statement by student that marks a specific and identifiable break from one topic of discussion to another | I have a question. This might be more tailored to our class but they have a lot of references. Are we supposed to be referencing a lot of other scientists, and theories, and then what kind of format, is this APA? (Immediately follows discussion of similarities and differences between discussion, analysis and conclusions.) |
| Negotiated Understanding | A group of students begins with several conflicting ideas and negotiates a common understanding | Discussion about whether title should be summary, statement, question or hypothesis ends with agreement that title should be a statement that "gives insight into your argument. |
| Distinguish Structures | Features that appear to be redundant have distinctions clarified | Discussion that begins with suggestion to combine discussion and conclusion and ends with clear operational distinction between discussion, analysis, and conclusion. |
| Instructor Prior Knowledge | Instructor uses prior knowledge to elaborate on an issue raised by students | So there might be anomalies in your data and . . . you need to explain why. [References a different course] really nice hyperbola except for the last two points . . . so yeah, if there's some things that don't fit, can you understand why? |
| Create Example | An illustrative example of a point is created | Like, not just "Analysis of Car Crashes" even though that's what I titled my paper. But . . . "The effect of whatever variable you're using on car crashes." |
| Functional Assessment | Applies an agreed-upon standard to evaluate a text | Use of a grading rubric |
| Positive Affect | Expresses positive emotion | The rubric for writing a scientific paper was the most clear rubric I've seen in my opinion. |
| Negative Affect | Expresses negative emotion | this one seemed like it was written by a middle schooler . . . You know, really, like, I was like, it was so bad to me |